\documentclass{appolb}
\usepackage{graphicx}
\newcommand{\eq}[1]{\begin{eqnarray}#1\end{eqnarray}}
\newcommand{\lsim}{\raisebox{-0.13cm}{~\shortstack{$<$ \\[-0.07cm] $\sim$}}~}
\newcommand{\gsim}{\raisebox{-0.13cm}{~\shortstack{$>$ \\[-0.07cm] $\sim$}}~}
\newcommand{\ii}{{\mathrm i}}

\begin{document}
\title{THE FULL LEPTON FLAVOR OF LITTLE HIGGS%
\thanks{Presented by J.I.~Illana at the XLIII International Conference of Theoretical Physics ``Matter to the Deepest'', Katowice, Poland, September 1-6, 2019.}%
}
\author{J.I.~Illana, 
        J.M.~P\'erez-Poyatos 
\address{CAFPE and Departamento de F{\'\i}sica Te\'orica y del Cosmos, \\         
         Universidad de Granada, E-18071 Granada, Spain}
}
\maketitle
\begin{abstract}
The little Higgs model with T-parity, compatible with electroweak precision constraints, introduces new flavor-mixing sources some of which had been ignored until recently. They are reviewed here, showing that their influence does not only enrich the phenomenology of flavor-changing processes but is also needed to render finite one-loop amplitudes.
\end{abstract}

\section{Introduction}

The Standard Model (SM) suffers from a hierarchy problem: the Higgs mass is of order $v$,  the electroweak scale, despite it receives quadratic loop corrections of the order of the theory cutoff, as large as the Planck scale if we take the model seriously up to the scale where gravity sets in. There are several attempts to solve this fine-tuning problem. In supersymmetric extensions, for instance, these corrections are cancelled by the contributions of superpartners.

Little Higgs (LH) models \cite{ArkaniHamed:2001ca} are tailored to solve the hierarchy problem in a very different way. There the Higgs is a pseudo-Goldstone boson of an approximate global symmetry spontaneously broken at a scale $f$, of the order of the TeV. The low energy degrees of freedom are described by a nonlinear sigma model, an effective theory valid below a cutoff $\Lambda\sim 4\pi f$ (order of $10$ TeV). As a consequence, the one-loop corrections to the Higgs
\eq{
\Delta M^2_h\sim\left\{y_t^2,g^2,\lambda^2\right\} \frac{\Lambda^2}{16\pi^2}\lsim (1\mbox{ TeV})^2,
}
from top, gauge and Higgs loops, are not fine-tuned. Ultraviolet completion, unknown but not needed, is required only for physics above $\Lambda$, beyond current experimental verification. The global symmetry is also explicitly broken by gauge and Yukawa interactions, giving the Higgs a mass and non-derivative interactions while preserving the cancellation of one-loop quadratic corrections thanks to the `collective symmetry breaking': the symmetry is broken only when two or more couplings are non-vanishing simultaneously.\footnote{%
By a different mechanism, low-scale supersymmetry is not exact either, but softly broken, still preserving the cancellation of quadratic corrections.}

As a byproduct, LH models introduce extra scalars, fermions and gauge bosons that bring new flavor mixing sources leading to enhanced decay rates of lepton flavor changing processes. This is the object of present talk.

\section{The littlest Higgs with T parity}

There are several types of LH models \cite{Perelstein:2005}. We focus here on the {\em littlest} Higgs \cite{ArkaniHamed:2002qy}, where the global symmetry is $SU(5)$, that breaks spontaneously to $SO(5)$ when the symmetric $5\times5$ scalar field tensor $\Sigma(x)$ of a nonlinear sigma model acquires a vacuum expectation value $\Sigma_0$ at a scale $f$. The 14 broken generators $X^a$ expand the Goldstone fields $\Pi(x)=\pi^a(x)X^a$, $\Sigma={\rm e}^{2\ii\Pi/f}\Sigma_0$. Only a subgroup $[SU(2)\times U(1)]_1\times[SU(2)\times U(1)]_2\subset SU(5)$ is gauged. The combinations of $SU(2)_i$ and $U(1)_i$ generators $Q_1^a+Q_2^a$, $Y_1+Y_2$ are unbroken when $\langle\Sigma\rangle=\Sigma_0$, so the 4 associated gauge bosons ($\gamma$, $Z$, $W^+$, $W^-$) remain massless, while the other 4 gauge bosons ($A_H$, $Z_H$, $W_H^+$, $W_H^-$), associated to the broken orthogonal combinations, get a mass proportional to $f$, absorbing 4 would-be-Goldstone bosons. The remaining 10 Goldstone bosons consist of a complex $SU(2)$ doublet $H$ and a complex triplet $\Phi$. In a second stage, 3 would-be-Goldstone bosons in $H$ are absorbed when 
the 3 standard weak gauge bosons ($Z$, $W^+$ and $W^-$) get a mass proportional to $v$ by the electroweak spontaneous symmetry breaking (EWSB). We are left with 7 physical pseudo-Goldstone bosons of the global $SU(5)$ symmetry: the Higgs field $h$, and 6 complex scalars in $\Phi$ ($\Phi^{\pm\pm}$, $\Phi^\pm$, $\Phi^0$, $\Phi^P$).

The presence of new particles at the TeV scale introduces significant loop corrections to electroweak observables, generating tension with precision tests. To alleviate it \cite{Hubisz:2005tx}, an additional discrete symmetry is introduced \cite{Cheng:2003ju}, the T-parity, under which the SM fields are even and (most of) the new fields are odd.\footnote{%
The T-parity in LH has the same purpose as the R-parity in supersymmetry.}

In the littlest Higgs model with T-parity (LHT) the gauge-scalar interactions of the fields described above are easy to implement by inserting the corresponding covariant derivatives. But the fermion sector is less straightforward. Regarding leptons, one introduces two left-handed $SU(2)$ doublets, $l_{1L}$ and $l_{2L}$, in incomplete $SU(5)$ multiplets ($\Psi_1$, $\Psi_2$), and two more, $l_{HR}$ and $\tilde\psi_R$, together with a singlet, $\chi_R$, in a complete $SO(5)$ multiplet ($\Psi_R$), 
\eq{
\Psi_1=\left(\begin{array}{c}-\ii\sigma^2 l_{1L}\\0\\0\end{array}\right), \quad
\Psi_2=\left(\begin{array}{c}0\\0\\-\ii\sigma^2 l_{2L}\end{array}\right), \quad
\Psi_R=\left(\begin{array}{c}\tilde\psi_R\\\chi_R\\-\ii\sigma^2 l_{HR}\end{array}\right).
}
The T-parity assignment is such that the combination $l_L=(l_{1L}-l_{2L})/\sqrt{2}$ is even, and identified with the SM left-handed lepton doublet ($\nu_L$, $\ell_L$), while the combination $l_{HL}=(l_{1L}+l_{2L})/\sqrt{2}$ is odd. The latter pairs with $l_{HR}$, also T-odd, to form a heavy Dirac doublet of {\em mirror} leptons ($\nu_H$, $\ell_H$). The fields $\tilde\psi_R$ (T-odd) and $\chi_R$ (T-even) are usually ignored assuming they are heavy and decouple in loop corrections. However, this is not always the case, and indeed the {\em mirror partners} in $\tilde\psi_R\equiv-\ii\sigma^2\tilde l^c_L$ play an important role in some lepton flavor changing processes as will be shown below.

In order to provide fermions with masses, several Yukawa interactions are introduced preserving gauge invariance and T parity. After the spontaneous breaking of the $SU(5)$ symmetry, the mirror leptons acquire a mass of order $f$ proportional to a coupling $\kappa$. The SM (charged) leptons, whose right-handed components are $SU(5)$ singlets, get masses of order $v$ times a coupling $\lambda_\ell$ by the EWSB. Because there is no other left-handed doublet to pair with the right-handed mirror partners, they cannot get their mass from Yukawa couplings, so one has to add another incomplete representation of $SO(5)$, $\Psi_L=(\tilde\psi_L,0,0)$ with $\tilde\psi_L\equiv-\ii\sigma^2\tilde l^c_R$, and include a generic mass term $M$ for them in the Lagrangian. Therefore, fermions masses come from 
\eq{
{\cal L} \supset -\frac{\lambda_\ell}{\sqrt{2}}v\;\overline{\ell_L}\ell_R
                 -\sqrt{2}\kappa f\;\overline{l_{HL}}l_{HR}
                -M\;\overline{\tilde l_L}\tilde l_R + {\rm h.c.}
}

\section{New sources of flavor mixing}

The T-parity is exact, so the SM fermions (T-even) do not mix with the T-odd fermions. Since there are several fermion families, $\lambda_\ell$, $\kappa$ and $M$ become matrices in flavor space. Their diagonalization leads to mass eigenstates and flavor mixings. Two mixing matrices are observable: $V$, expressing the misalignment between $l_H$ and $l$; and $W$, showing the misalignment between $\tilde l^c$ and $l_H$. The phenomenology of the former has been extensively studied (see \cite{Blanke:2006sb,Buras:2006wk,Blanke:2006eb,Goto:2008fj,delAguila:2008zu,delAguila:2010nv} and references therein). But the relevance of the latter and its implications have been emphasized only recently \cite{delAguila:2017ugt,delAguila:2019htj}. 

To simplify the discussion we will consider two-family mixing, parameterized by just two mixing angles ($\theta_V$ and $\theta_W$, that must not to be confused with the weak mixing angle),
\eq{
V=\left(\begin{array}{cc}\cos\theta_V & \sin\theta_V \\
                        -\sin\theta_V & \cos\theta_V \end{array}\right),\quad
W=\left(\begin{array}{cc}\cos\theta_W & \sin\theta_W \\
                        -\sin\theta_W & \cos\theta_W \end{array}\right),\quad
}
and the masses of mirror leptons and mirror partners, that can be traded for two average masses ($\overline{m}_{\ell_H}$, $\overline{m}_{\tilde\nu^c}$) and two mass splittings ($\delta_{\ell_H}$, $\delta_{\tilde\nu^c}$),
\eq{
\overline{m} &= \sqrt{m_1m_2}, \quad
\delta = (m_2^2-m_1^2)/\overline{m}^2.
}
The genuine lepton-flavor violating (LFV) effects from this LH extension of the SM vanish in case of zero mixings (full alignment of standard with mirror and mirror partner leptons) or zero splittings (mirror and mirror partner masses are degenerate).

\section{Lepton flavor changing processes}

Let us explore the impact of these new LFV sources on the phenomenology by analyzing the one-loop contributions of T-odd particles to the lepton flavor changing processes $h \to \bar\ell\ell'$, $Z\to\bar\ell\ell'$, $\ell\to\ell'\gamma$, $\ell\to\ell'\ell'\bar\ell',\ell'\ell''\bar\ell'',\ell'\ell'\bar\ell''$ and $\mu \to e$ conversion in nuclei. The calculation \cite{delAguila:2008zu,delAguila:2010nv,delAguila:2017ugt,delAguila:2019htj} is based on the careful derivation of all Feynman rules needed, after the expansion of the Lagrangian in powers of $v/f$, and the meticulous analytical decomposition in the standard basis of tensor integrals of the one-loop diagrams involved. The leading order of the resulting amplitudes is $(v/f)^2$ in all cases.

The amplitude for Higgs decays has the following Lorentz structure,
\eq{
{\cal M}(h\to\bar\ell\ell') = \overline{u}(p_2)\left(
\frac{m_{\ell'}}{v} c_L P_L + \frac{m_{\ell}}{v} c_R P_R \right)v(p_1),
}
where the coefficients $c_{L(R)}$ are functions of the masses and mixings of the T-odd particles running in the loop,
\eq{
c_{L(R)} &=& \frac{g^2}{16\pi^2}\frac{v^2}{f^2}\Big\{
 \sum_i V_{\ell i}^\dagger V_{i\ell'}\; {\cal F}_h(m_{\ell_{Hi}},\dots) 
\nonumber \\
&+&\sum_{i,j,k} V_{\ell i}^\dagger \frac{m_{\ell_{Hi}}}{m_{W_H}} 
              W_{ij}^\dagger W_{jk} \frac{m_{\ell_{Hk}}}{m_{W_H}} 
              V_{k\ell'} \; {\cal G}_h(m_{\tilde\nu^c_j},m_{\ell_{Hk(i)}},\dots)
\Big\}.
}
The second line contains non-decoupling contributions from mirror partner leptons. In fact, for the case of two-family mixing, in the limit $m_{\tilde\nu^c_j}\to\infty$,
\eq{
c_{L(R)}(\theta_W=\delta_{\tilde\nu^c}=0)&\sim& \frac{g^2}{16\pi^2}\frac{v^2}{f^2}\sin2\theta_V\frac{\overline{m}^2_{\ell_H}}{M^2_{W_H}}\delta_{\ell_H},
\\
c_{L(R)}(\theta_V=\delta_{\ell_H}=0)&\sim& \frac{g^2}{16\pi^2}\frac{v^2}{f^2}\sin2\theta_W\frac{\overline{m}^2_{\ell_H}}{M^2_{W_H}}\ln\frac{m^2_{\tilde\nu^c_2}}{m^2_{\tilde\nu^c_1}}.
}
 
The effective vertices for $Z\to\bar\ell\ell'$ and $\ell\to\ell'\gamma$ have identical structure,
\eq{
\Gamma^\mu_V = e\left[F_L^V(q^2)\gamma^\mu P_L + \ii F_M^V(q^2)(1+\gamma_5)\sigma_{\mu\nu}q_\nu\right],
\label{eq:10}
}
where $q^2$ is the squared momentum of $V=\gamma,Z$, and we have used that, up to corrections of order $m_\ell^2/f^2$, in the LHT $F_R^V=0$ and $F_M^V=-\ii F_E^V$. Due to the $U(1)_{\rm em}$ gauge invariance, the form factor $F_L^\gamma$ vanishes for on-shell photons ($q^2=0$) since $\ell\to\ell'\gamma$ is a dipole transition. As a consequence, $F_L^\gamma(q^2)\sim q^2/(4\pi f)^2$ for small $|q^2|$, while $F_L^Z(0)\sim F_L^Z(M_Z^2)\sim v^2/(4\pi f)^2$. The dipole form factors are $F_M^{\gamma,Z}\sim m_\ell/(4\pi f)^2$. 

The amplitudes for $\ell\to\ell'\ell'\bar\ell',\ell'\ell''\bar\ell'',\ell'\ell'\bar\ell''$
are obtained by the coherent sum of photon-penguin (${\cal M}_\gamma$), $Z$-penguin (${\cal M}_Z$) and box (${\cal M}_{\rm box}$) diagrams,
\eq{
{\cal M}_\gamma &=& \frac{e}{q^2}
\left[\overline{u}(p_1)\Gamma_\gamma^\mu u(p)\right]
\left[\overline{u}(p_2)\gamma_\mu v(p_3)\right] ,
\\
{\cal M}_Z &=& -\frac{e}{M_Z^2}
\left[\overline{u}(p_1)\Gamma_Z^\mu u(p)\right]
\left[\overline{u}(p_2)\gamma_\mu\left(Z_L P_L + Z_R P_R \right)v(p_3)\right] ,
\\
{\cal M}_{\rm box} &=& e^2 B_L 
\left[\overline{u}(p_1)\gamma^\mu P_L u(p)\right]
\left[\overline{u}(p_2)\gamma_\mu P_L v(p_3)\right] ,
}
where $Z_{L,R}$ are the tree-level couplings of a $Z$ to (two same-flavor) leptons, and crossed diagrams with $p_1$ and $p_2$ exchanged must be included as well, where appropriate. This amplitude involves the effective vertices in (\ref{eq:10}) whose form factors are evaluated in the limit $|q^2|\ll M_Z^2\ll f^2$. Then the dynamics is captured by the coefficients:
\eq{
A_L \equiv \lim_{q^2\to 0}\frac{F_L^\gamma(q^2)}{q^2}, \;\;
A_R \equiv \frac{2 F_M^\gamma(0)}{m_\ell}, \;\;
F_{LL(LR)} \equiv -\frac{F_L^Z(0)}{M_Z^2}Z_{L(R)}, \;\;
B_L.
}
All these coefficients are of order $(4\pi f)^{-2}=\Lambda^{-2}$. The $F_M^Z$ dipole form factors do not contribute here, since they are further suppressed by a factor $m_\ell^2/f^2$. The amplitude for $\mu\,{\rm N}\to e\,{\rm N}$ (conversion in nuclei) is described by the same $A_{L,R}$, $F_{LL,LR}$ and similar box form factors $B_L^q$, obtained by replacing one  of the lepton currents by a current of quarks $q=u,d$.  

The various one-loop contributions to dipole and box form factors are finite. However, every $c_{L,R}$, $F_L^\gamma$ and $F_L^Z$ contain ultraviolet-divergent contributions that cancel each other only when {\em all} possible gauge and T invariant interactions are included, making the LHT model {\em predictable} at leading order in $v/f$. In particular, loop diagrams with mirror partner leptons $\tilde l^c$ accompanied by $\Phi$ scalars are needed to yield a finite $c_{L,R}$ and hence finite amplitudes for LFV Higgs decays. This fact is related to the non-decoupling of mirror partner leptons in loops for $h$ decay amplitudes. In contrast, if taken sufficiently heavy, mirror partner lepton contributions can be ignored in the other (gauge-mediated) processes. 

In order to estimate the typical size of the LFV effects induced by the LHT, we take the following set of ``natural'' {\it reference values} for the input parameters \cite{delAguila:2019htj}: $f=1.5$~TeV, $\overline{m}_{\ell_H}=\overline{m}_{\tilde\nu^c}=1$~TeV, $\overline{m}_{d_H}=\overline{m}_{\tilde u^c}=2$~TeV, $\delta_{\ell_H}=\delta_{\tilde\nu^c}=1$, $\sin 2\theta_V=\sin 2\theta_W=1$, assuming for simplicity that the flavor change occurs only between the two families involved in the transition. These values are compatible with direct searches at the LHC \cite{Dercks:2018hgz}. As one can see in Tables~\ref{tab:1} and \ref{tab:2}, the $\mu$ to $e$ transitions are very constrained by current experimental bounds, particularly from $\mu\to e\gamma$ and $\mu\to e$ conversion in nuclei, whereas the transitions involving the third family do not set restrictions.  The processes with double lepton flavor change, $\tau\to\mu\mu\bar{e}$ and $\tau\to ee\bar{\mu}$, would be forbidden under the assumption of just two-family mixing. LFV $Z$ and Higgs decays are far from being at reach in this model.

\begin{table}
\centerline{%
\begin{tabular}{|l|ll|}
\hline
& LHT & bounds \\
\hline\hline
${\cal B}(\mu\to e\gamma)$ & $4.3\cdot10^{-9}$ & $4.2\cdot10^{-13}$ \\
${\cal R}(\mu\,{\rm Au}\to e\,{\rm Au})$ & $3.8\cdot10^{-9}$ & $7.0\cdot10^{-12}$ \\
${\cal B}(\mu\to ee\bar{e})$ & $2.5\cdot10^{-11}$ & $1.0\cdot10^{-12}$ \\
${\cal B}(Z\to \mu \bar{e} + \bar{\mu} e)$ & $2.7\cdot10^{-12}$ & $7.3\cdot10^{-7}$ \\
${\cal B}(h\to \mu \bar{e} + \bar{\mu} e)$ & $1.2\cdot10^{-15}$ & $3.5\cdot10^{-4}$ \\
\hline
\end{tabular}}
\caption{Predictions for $\mu-e$ transitions in the LHT with reference inputs and current 90\% C.L. experimental bounds (95\% C.L. for $Z$ and $h$ decays). From \cite{delAguila:2019htj}. \label{tab:1}}
\end{table} 

\begin{table}
\centerline{%
\begin{tabular}{|l|ll||l|ll|}
\hline
& LHT & bounds &
& LHT & bounds \\
\hline\hline
$\tau\to  e\gamma$ & $7.3\cdot10^{-10}$ & $3.3\cdot10^{-8}$ &
$\tau\to\mu\gamma$ & $7.3\cdot10^{-10}$ & $4.4\cdot10^{-8}$ \\
$\tau\to e\mu\bar\mu$ & $2.2\cdot10^{-12}$ & $2.7\cdot10^{-8}$ &
$\tau\to \mu e\bar e$ & $8.2\cdot10^{-12}$ & $1.8\cdot10^{-8}$ \\
$\tau\to ee\bar{e}$ & $7.4\cdot10^{-12}$ & $2.7\cdot10^{-8}$ &
$\tau\to \mu\mu\bar{\mu}$ & $1.4\cdot10^{-12}$ & $2.1\cdot10^{-8}$ \\
$Z\to \tau e$ & $2.7\cdot10^{-12}$ & $9.8\cdot10^{-6}$ &
$Z\to \tau \mu$ & $2.7\cdot10^{-12}$ & $1.2\cdot10^{-5}$ \\
$h\to \tau e$ & $3.2\cdot10^{-13}$ & $5.2\cdot10^{-3}$ &
$h\to \tau \mu$ & $3.2\cdot10^{-13}$ & $2.5\cdot10^{-3}$ \\
\hline
\end{tabular}}
\caption{As in Table~\ref{tab:2} for transitions involving the third lepton family. \label{tab:2}}
\end{table}

\begin{figure}[htb]
\centerline{%
	    \includegraphics[scale=0.577]{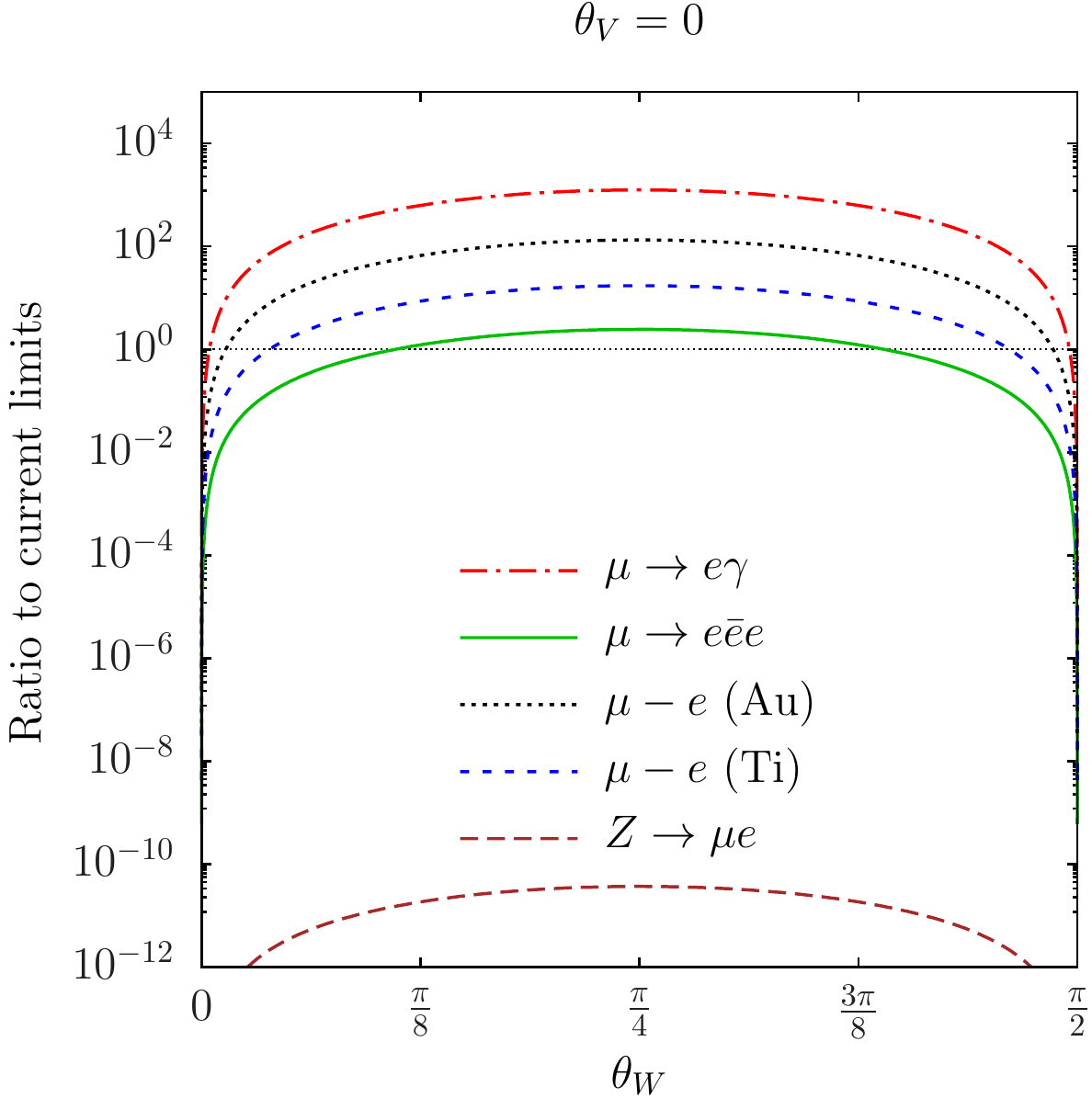}} 
\caption{$\mu$ to $e$ transitions as a function of the $\theta_W$ mixing angle. From \cite{delAguila:2019htj}. \label{Fig:1}}
\end{figure}

\begin{figure}[htb]
\centerline{%
		\includegraphics[scale=0.5]{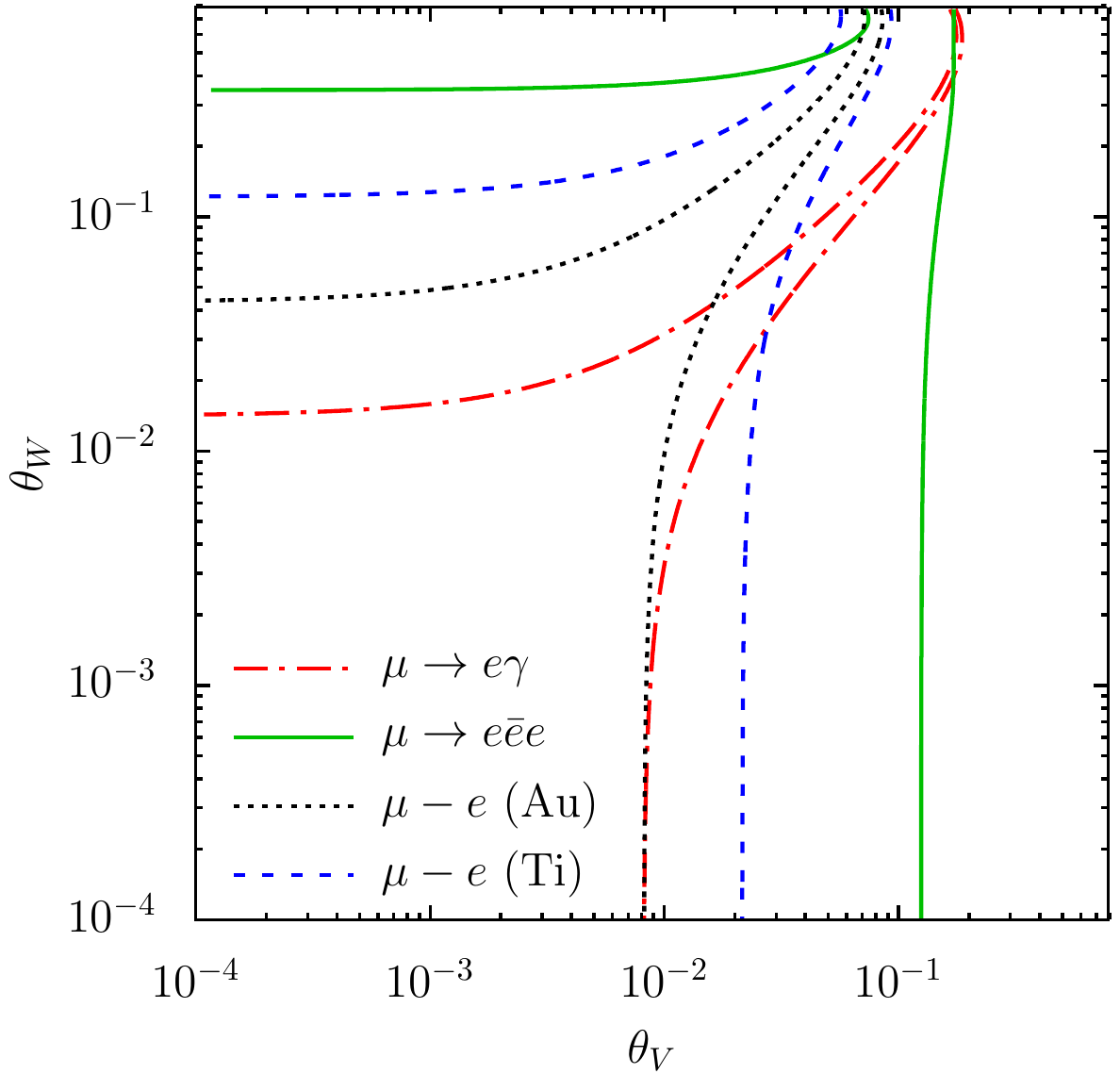}\hfill
		\includegraphics[scale=0.5]{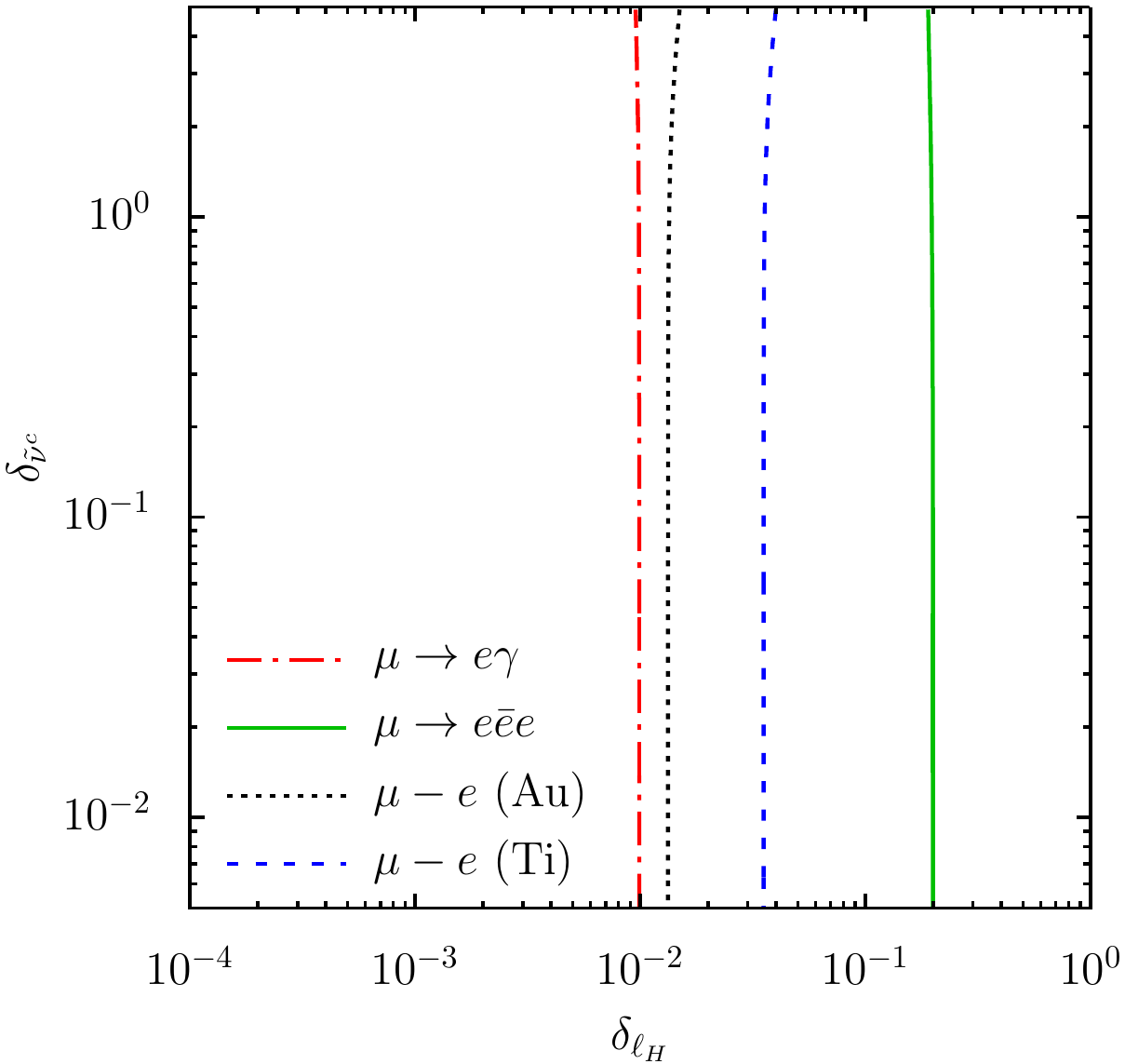}}
\caption{Contours saturating current bounds on $\mu$ to $e$ transitions. From \cite{delAguila:2019htj}. 
\label{Fig:2}}
\end{figure}

\begin{figure}[htb]
\centerline{%
		\includegraphics[scale=0.6]{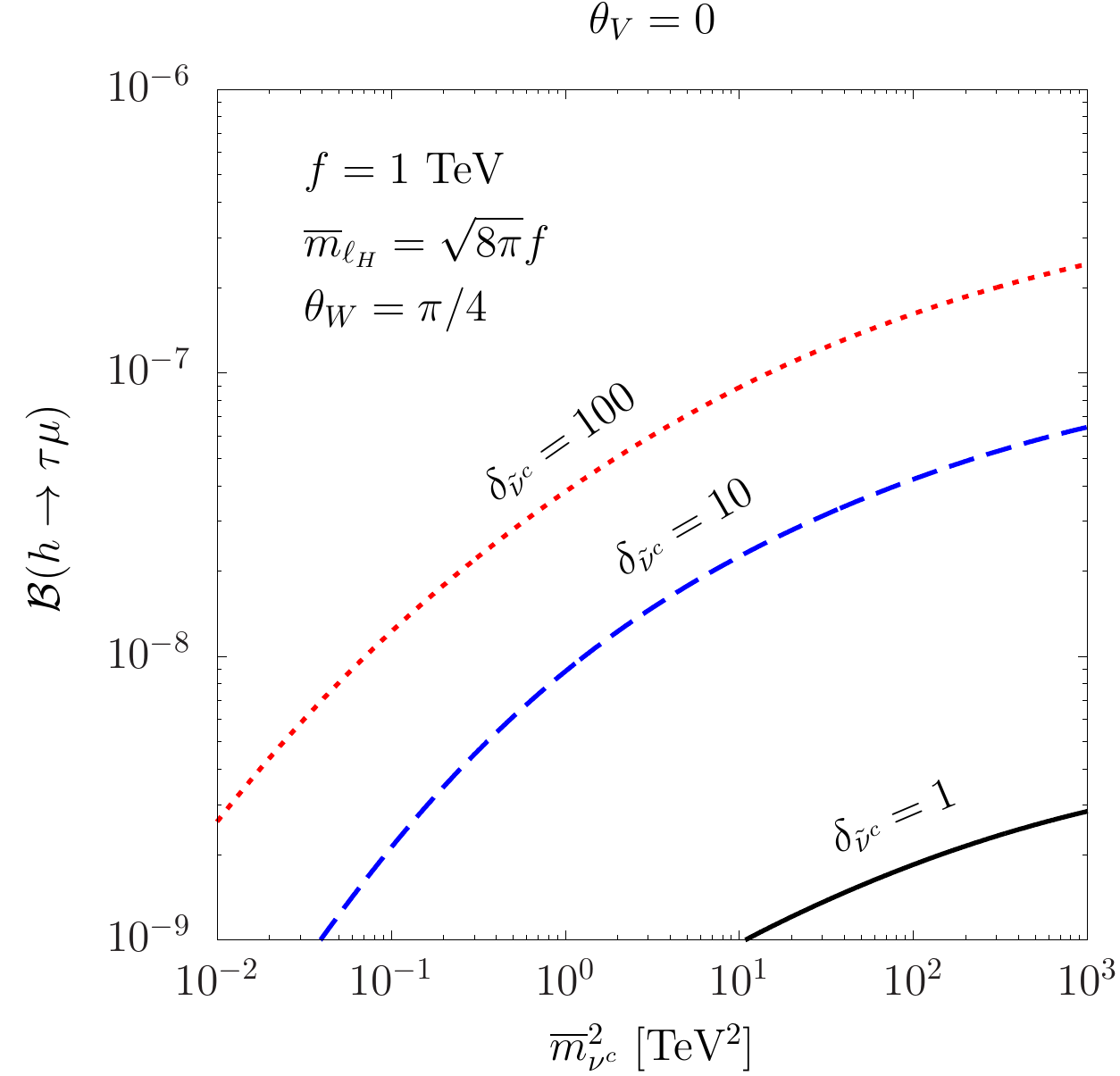}}
\caption{Non-decoupling behavior of ${\cal B}(h\to\tau\mu)$ with the mass of the mirror partners. \label{Fig:3}}
\end{figure}

The dependence of the various LFV observables on the parameters of the model is illustrated by performing scans over one or two of them with the others fixed to their reference values, unless otherwise stated. For instance, in Fig.~\ref{Fig:1} the rates of different $\mu$ to $e$ transitions normalized to their current limits as a function of the $\theta_W$ mixing angle for $\theta_V=0$ shows that $\mu\to e\gamma$ is the most sensitive, allowing only very small values of $\sin2\theta_W$. Fig.~\ref{Fig:2} shows contour lines in the plane of the two mixing angles (left) and of the two mass splittings (right) that give maximal rates for $\mu$ to $e$ transitions allowed by current bounds, forcing both mixings and the $\delta_{\ell_H}$ to be smaller than $10^{-2}$. If one would take $\theta_V=0$ then also $\delta_{\tilde\nu^c}$ would be constrained to be below $10^{-2}$. 
In Fig.~\ref{Fig:3} the non-decoupling of the mirror partner leptons in Higgs decays is apparent. We see that, taking a maximal value $\sim\sqrt{4\pi}$ for the $\kappa$ Yukawa couplings, compatible with perturbative unitarity, the branching ratio ${\cal B}(h\to\tau\mu)$ can be larger than $10^{-7}$ for heavy mirror partner leptons with large mass splittings. Note that for $\delta_{\tilde\nu^c}\gg 1$, $m^2_{\tilde\nu^c_1}\to \overline{m}^2_{\tilde\nu^c}/\delta_{\tilde\nu^c}$ and $m^2_{\tilde\nu^c_2}\to \overline{m}^2_{\tilde\nu^c}\times\delta_{\tilde\nu^c}$.

It is interesting to remark that when the contribution of the photon dipole form factor $A_R$ dominates the three-body $\tau$ decay, the ratio to the corresponding radiative $\tau$ decay is fixed by kinematics,
\eq{
\frac{{\cal B}(\tau\to\mu\mu\bar{\mu})}{{\cal B}(\tau\to\mu\gamma)}
\approx2\times10^{-3}, \quad
\frac{{\cal B}(\tau\to\mu e\bar{e})}{{\cal B}(\tau\to\mu\gamma)}
\approx10^{-2}, \quad
\frac{{\cal B}(\tau\to\mu e\bar{e})}{{\cal B}(\tau\to\mu\mu\bar{\mu})}\approx 5.\quad
}
The dipole dominance occurs in SUSY, but not in the LHT \cite{Blanke:2007db,Blanke:2009am} {\it if} the contributions of mirror partner leptons are negligible, namely when they are heavy and decouple. In this case ${\cal B}(\tau\to\mu e\bar{e})\approx{\cal B}(\tau\to\mu\mu\bar{\mu})$ in contrast to SUSY predictions. However, if the masses of $\ell_H$ and $\tilde l^c$ are of the same order, the LHT contributions to dipole form factor dominate as in SUSY, and both models behave the same.

The current experimental bounds on $\mu$ to $e$ transitions translate into generic limits on the scale $f$ and the mixing angles of the LHT model. They are summarized in Table~\ref{tab:3} (left columns), together with the expected limits (right columns) that would be derived in the absence of any signal in future phases of those experiments. For a natural value of $f=1.5$~TeV, the mixings are already restricted to be of ${\cal O}(10^{-2})$ by $\mu\to e\gamma$. These constraints would be improved by two orders of magnitude from $\mu\,{\rm Al} \to e\,{\rm Al}$ in the future. 

\begin{table}
\centerline{%
\begin{tabular}{|lc|cc|cc|}
\hline
\multicolumn{2}{|c|}{} & \multicolumn{2}{c|}{[$\theta_V = \theta_W = \pi/4$]} & \multicolumn{2}{c|}{[$f=1.5$~TeV]} \\
Process & Experiment &
\multicolumn{2}{c|}{$f$ [TeV] $>$}  & \multicolumn{2}{c|}{$\theta_{V,W}/10^{-2} <$}
\\ \hline\hline
$\mu \to e\gamma$ & [MEG]
&  {\boldmath$15$}  & {$27-71$} & \, {\boldmath$1$}  & {$0.3 - 0.04 $} 
\\ 
$\mu \to ee\overline{e}$ & [Mu3e] &  {$3.4$}  & {$13-34$} & {$20$}  & {$1-0.3 $} 
\\ 
$\mu\,{\rm Al} \to e\,{\rm Al}$ & [Mu2e] 
&  {$8.4$}  & {\boldmath$84-150$} &\, {$3$}  & {\boldmath$0.03 - 0.01$} 
\\   
$\mu\,{\rm Al} \to e\,{\rm Al}$ & [COMET]
&  {$8.4$}  & {$27-84$} & \, {$3$}  & {$0.3 - 0.03 $} 
\\  
\hline
\end{tabular}}
\caption{Present and future limits on LHT parameters. From \cite{delAguila:2019htj}. \label{tab:3}}
\end{table}

\begin{figure}[htb]
\centerline{%
		\includegraphics[scale=0.6]{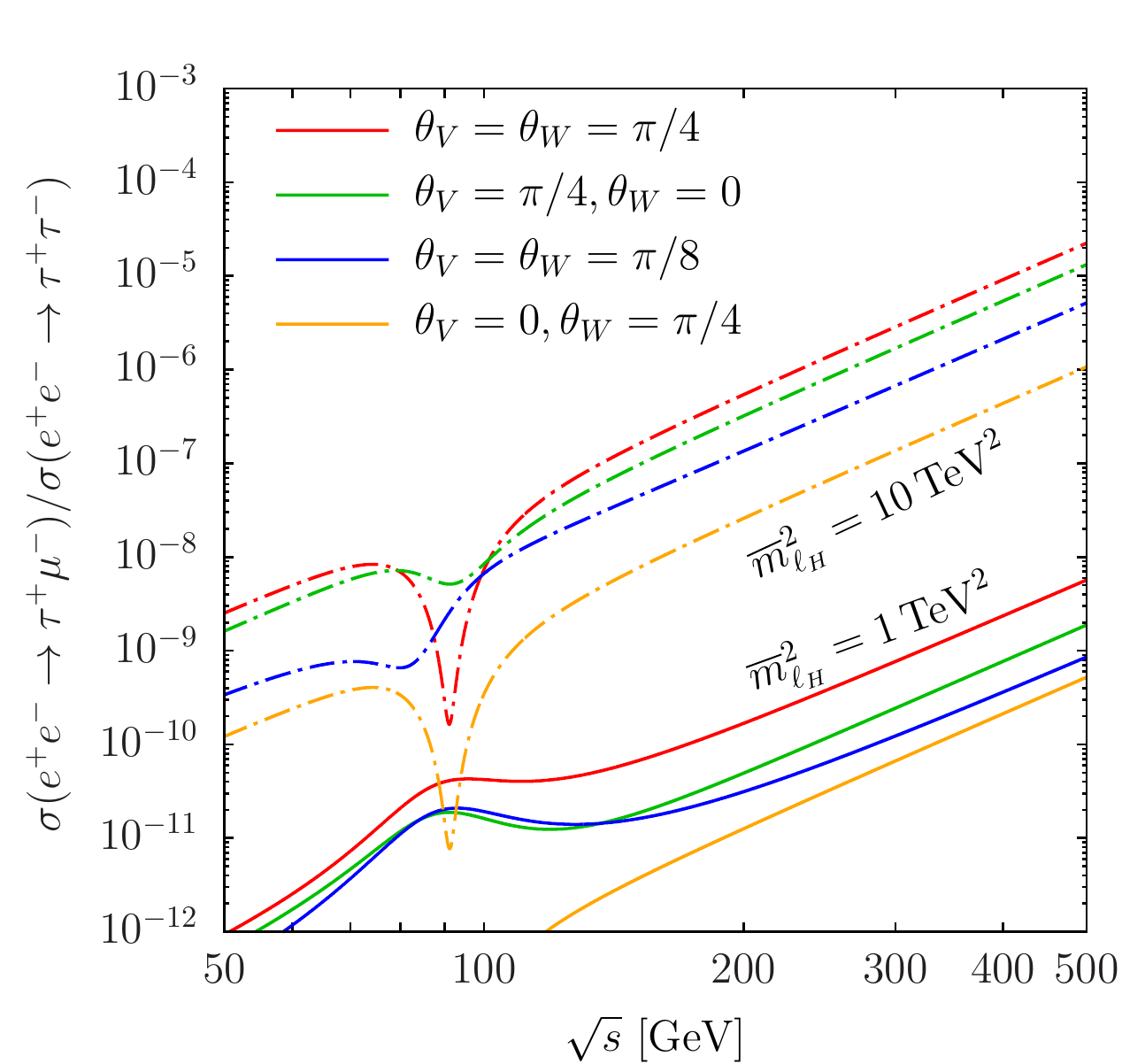}}
\caption{Ratio of $\tau^+ e^-$ to $\tau^+\tau^-$ production as a function of the center of mass energy in $e^+e^-$ collisions. From \cite{delAguila:2019htj}. \label{Fig:4}}
\end{figure}

Regarding $\tau$ to $e$ and $\tau$ to $\mu$ transitions, the current bounds on LFV $\tau$ decays (see Table~\ref{tab:2}) will be improved by one or two orders of magnitude at Belle II and LHCb, starting to constrain the parameter space of the third family. The situation would be even more promising at future lepton colliders \cite{Dam:2018rfz}. Thanks to the growth with $s$ of the cross-section $e^+e^-\to\tau^\pm\mu^\mp$, due to the box contributions, one may find enough LFV production for high energies even if it is not observed at the $Z$ peak (Fig.~\ref{Fig:4}). Since one expects $10^{10}$ $\tau^+\tau^-$ pairs for an integrated luminosity of $10\,{\rm ab}^{-1}$, a good portion of parameter space would be probed. 

\section{Conclusions}

The one-loop predictions for flavor violating processes in the LHT model are ultraviolet-finite when {\it all} Goldstone interactions compatible with the gauge and T symmetries and {\it all} T-odd leptons are included.

Apart from negligible contributions of light neutrinos (suppressed by their tiny masses), lepton flavor mixing in this model is due to the misalignment of mirror leptons ($\nu_H$, $\ell_H$) with the SM charged leptons, and {\it also} with mirror partners ($\tilde\nu^c$, $\tilde\ell^c$). The former is a well known source of LFV, and the latter provides additional contributions that cannot be ignored because they are needed to make LFV Higgs decay amplitudes finite. Actually, mirror partner leptons do not decouple in $h \to \overline{\ell}\ell'$ where they contribute at one loop accompanied by the
scalar triplet $\Phi$ at leading order in $v/f$. This is in contrast to gauge mediated LFV processes ($Z\to \bar\ell\ell'$, $\ell\to\ell'\gamma$, $\mu\,{\rm N}\to e\,{\rm N}$, $\ell\to\ell'\ell''\bar\ell'''$, etc.), where the contribution of $\Phi$ is subleading and the mirror lepton partners do decouple if they are heavy. The masses of mirror leptons and mirror partners have a very different origin, but, if taken of the same order, their contributions to all LFV processes are of similar size.\footnote{%
For implications of the field $\chi_R$ on neutrino masses and LFV see \cite{delAguila:2019mvp}.
}

To summarize, flavor provides complementary constraints to LH models, particularly from $\mu-e$ transitions. Current experiments already require a small misalignment between leptons, mirror leptons and their mirror partners, with mixings and mass splittings smaller than $10^{-2}$, or a high scale $f\gsim 10$~TeV, reintroducing the unnaturalness that little Higgs came to solve.

Work supported in part by the Spanish Ministry of Science, Innovation and Universities, under grants FPA2016-78220-C3-1,2,3-P (fondos FEDER), and Junta de Andaluc{\'\i}a, grants FQM~101 and SOMM17/6104/UGR. J.I.I. wishes to thank F.~del~Aguila,
Ll.~Ametller, M.D.~Jenkins, T.~Hahn, J.~Santiago, P.~Talavera and R.~Vega-Morales for a fruitful collaboration and 
the Conference Organizers for the very pleasant atmosphere of the meeting.

\end{document}